# Enhanced Hyperspectral Nano-FTIR Imaging by Neural-Network-Based Processing of Multiple Signal Harmonics

Pablo Calvo-Barlés[1,2], Sergio G. Rodrigo[1,3], Luis Martín-Moreno[1,2*], Martin Schnell[4,5,6], Rainer Hillenbrand[4,6,7*]

[1] *Instituto de Nanociencia y Materiales de Aragón (INMA), CSIC-Universidad de Zaragoza, Zaragoza 50009, Spain*
[2] *Departamento de Física de la Materia Condensada, Universidad de Zaragoza, Zaragoza 50009, Spain*
[3] *Departamento de Física Aplicada, Universidad de Zaragoza, Zaragoza 50009, Spain*
[4] *CIC nanoGUNE BRTA, 20018 Donostia-San Sebastián, Spain*
[5] *Donostia International Physics Center, Donostia-San Sebastian 20018, Spain*
[6] *IKERBASQUE, Basque Foundation for Science, 48013 Bilbao, Spain*
[7] *Department of Electricity and Electronics, FCT-ZTF, University of the Basque Country EHU, Barrio Sarriena z/g, Leioa 48940, Spain.*

** Corresponding authors*

*Email: lmm@unizar.es (LMM), r.hillenbrand@nanogune.eu (RH)*

## Abstract
Nano-FTIR spectroscopy combines infrared spectroscopy with nanometer-scale spatial resolution and enables hyperspectral nanoimaging. However, quantitative measurements typically rely on higher-harmonic near-field signals, which are largely background-free but suffer from low signal-to-noise ratio (SNR). Lower harmonics provide substantially higher SNR but contain significant background contributions that distort the near-field response and hinder quantitative analysis. Here, we introduce a cross-harmonic neural network (CHNN) that exploits the complementary information encoded in multiple harmonic channels to denoise third-harmonic hyperspectral data. Trained directly on experimental nano-FTIR datasets, the CHNN uses first- and second-harmonic spectra to improve the SNR of the third-harmonic response. Applied to a PMMA/PS polymer blend, it substantially reduces noise while preserving spectral features and nanoscale spatial contrast. Compared with conventional single-harmonic denoising using an autoencoder, the CHNN achieves superior noise suppression. These results demonstrate the potential of lower-harmonic signals for improving hyperspectral nano-FTIR imaging without increasing acquisition time.

## Introduction

Scattering-type scanning near-field optical microscopy (s-SNOM) and nano-FTIR spectroscopy are nanoscale optical imaging and spectroscopy techniques based on atomic force microscopy (AFM) [1–4]. In these techniques, the AFM tip is illuminated by light and acts as a nanoscale antenna, concentrating the incident radiation into a highly localized optical near-field at the tip apex [5,6]. This near-field interacts with the sample and modifies the optical field scattered by the tip. The scattered field therefore contains information about the local optical properties of the sample. By recording the scattered field while raster-scanning the tip across the sample surface, optical images can be obtained with a spatial resolution determined by the tip apex radius rather than by the wavelength of light. Applicable from the visible and infrared to the terahertz spectral range [3,4,7], s-SNOM and nano-FTIR are widely used for nanoscale chemical imaging [8–12], materials characterization [2,4,8], semiconductor metrology [13–15], biological [16–21] and polymer analysis [22–26], and the investigation of plasmonic [27–31], phononic [32–34], and other polaritonic excitations [35,36]. In nano-FTIR spectroscopy, the tip is positioned at a fixed location while being illuminated with broadband infrared radiation [37–44]. The tip-scattered field is recorded with an asymmetric Fourier-transform interferometer by scanning the position of the reference mirror, yielding an interferogram whose Fourier transform provides amplitude- and phase-resolved infrared spectra of the local near-field interaction. In this way, nano-FTIR combines broadband vibrational spectroscopy with nanometer-scale spatial resolution.

The tip-scattered field contains both the desired near-field contribution, originating from the localized interaction between the optical near-field at the tip apex and the sample, and an undesired background contribution arising from scattering by the tip shaft, sample surface, and other optical components. Separating these contributions has long been a central challenge in s-SNOM and nano-FTIR [45–47]. This challenge is addressed by oscillating the AFM tip vertically at its mechanical resonance frequency, $\Omega$, detecting the tip-scattered radiation interferometrically, and demodulating the detector signal at harmonics of the tip oscillation frequency, $n\Omega$ [48]. Because the near-field interaction depends strongly and nonlinearly on the tip-sample distance, the near-field contribution generates significant higher-harmonic components, whereas the background contribution is concentrated predominantly at lower harmonics. Consequently, the near-field amplitude and phase signals become increasingly dominated by genuine near-field information with increasing harmonic order $n$ [45,46,49].

Higher demodulation orders, typically $n = 3$ and 4, provide signals that are largely background-free but are challenged by low signal-to-noise ratio (SNR). This limitation becomes particularly severe in hyperspectral nano-FTIR imaging, where a complete infrared spectrum is acquired at every image pixel, yielding hyperspectral data cubes with two spatial dimensions and one spectral dimension. Because quantitative hyperspectral measurements generally rely on the nearly background-free higher harmonics, acquisition times can easily reach several hours even for datasets comprising only a few thousand pixels (e.g., 86×62 pixels in ref. [25]). In contrast, the lower harmonics, particularly $n = 1$ and 2, offer substantially higher SNR but retain significant background contributions, whose coherent interference with the near-field signal can distort optical resonances by inducing asymmetric Fano profiles, shifting peak frequencies, and causing phase-dependent amplitude modulations that prevent reliable quantitative analysis.

Here, we introduce a Deep Learning (DL) approach that exploits the complementary information encoded in hyperspectral data cubes acquired at different harmonic orders. By combining the high-SNR information available in the first and second harmonics with the near-field purity of the third harmonic, our trained models reconstruct background-free higher-order hyperspectral data cubes with substantially enhanced signal-to-noise ratio. Unlike conventional denoising approaches that operate on a single harmonic hyperspectral data cube, our method leverages the physical correlations between multiple harmonic channels, enabling recovery of weak near-field signals while preserving spectroscopic fidelity.

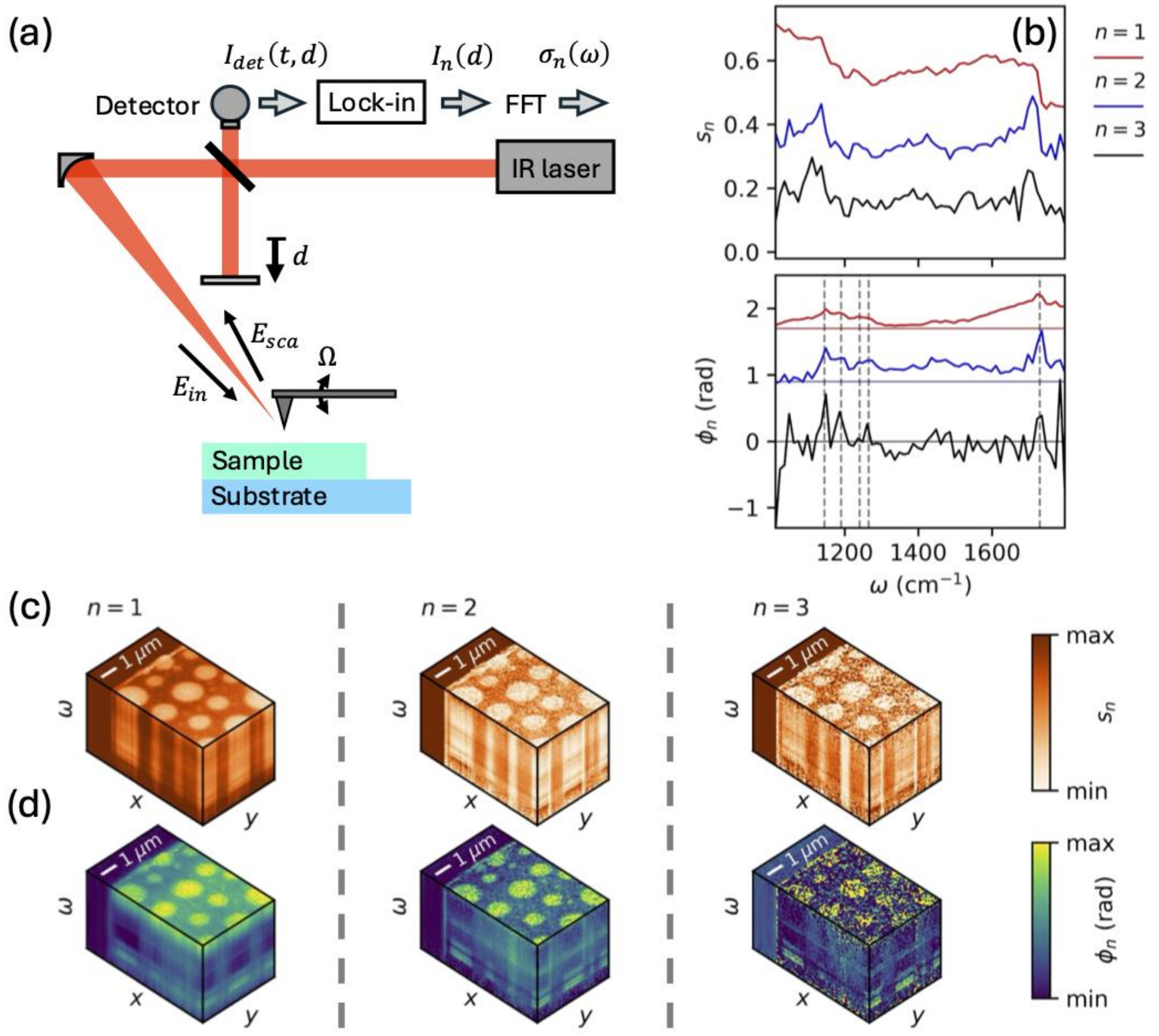

**Figure 1: Hyperspectral infrared nanoimaging based on nano-FTIR spectroscopy. (a)** Schematic representation of the experimental setup. **(b)** Amplitude and phase spectra, $s_n(\mathbf{r},\omega)$ and $\phi_n(\mathbf{r},\omega)$, at a representative fixed position $\mathbf{r}$ within a PMMA domain for the first three harmonics ($n = 1,2,3$), shown in red, blue, and black, respectively. Dashed vertical lines in the phase spectrum mark the PMMA vibrational resonances reported in ref. *[39]*. **(c,d)** Hyperspectral amplitude and phase data, $s_n(\mathbf{r},\omega)$ and $\phi_n(\mathbf{r},\omega)$, of size $N_x \times N_y \times N_\omega = 80 \times 49 \times 64$ recorded on a PMMA/PS polymer blend sample. The amplitude datacubes are displayed using the numerical ranges (min, max) = (0.3, 0.7), (0.25, 0.5), and (0.07, 0.3) for $n = 1,2,3$, respectively. The phase datacubes use independently adjusted ranges of (min, max) = (0.0, 0.5) rad, (0.0,0.7) rad, and (−0.2, 0.6) rad for $n = 1,2,3$, respectively.

The nano-FTIR measurement scheme is illustrated in Fig. 1(a). A broadband infrared laser illuminates a metallized AFM tip positioned above the sample surface. The tip acts as an infrared antenna, generating a highly localized near field at its apex. Interaction of this near field with the sample modifies the infrared field scattered by the tip. The scattered field can be expressed as

$$E_{\mathrm{sca}} = E_{\mathrm{NF}} + E_{\mathrm{BG}}, \qquad (1)$$

where $E_{\mathrm{NF}}$ and $E_{\mathrm{BG}}$ denote the near-field and background contributions, respectively. To measure $E_{\mathrm{sca}}$, the tip-scattered field is analyzed with an asymmetric Michelson interferometer, where it interferes at the detector with a reference field reflected from a movable mirror at position $d$.

To isolate the near-field signal, the AFM tip is oscillated vertically at the tapping frequency $\Omega$. Owing to the strong nonlinear dependence of the near-field interaction on the tip-sample distance, $E_{\mathrm{NF}}$ generates substantial signal components at higher harmonics $n\Omega$, whereas $E_{\mathrm{BG}}$ contributes predominantly to the lower harmonics. The detector signal is therefore demodulated at harmonic order $n$, enhancing the relative contribution of the near-field signal. Recording the demodulated signal as a function of mirror position $d$ yields the interferogram $I_n(d)$. Fourier transform of $I_n(d)$ with respect to $d$ yields the complex-valued nano-FTIR spectrum denoted $\sigma_{n,\mathrm{sample}}(\omega)$.

For hyperspectral nano-FTIR imaging, this procedure is repeated at every image pixel $\mathbf{r} = (x, y)$, yielding the hyperspectral data cube $\sigma_{n,\mathrm{sample}}(\mathbf{r}, \omega)$. To eliminate the influence of the illumination spectrum and instrument response, a reference spectrum $\sigma_{n,\mathrm{ref}}(\omega)$ is acquired on an exposed region of the substrate. The referenced complex-valued nano-FTIR hyperspectral datacube is then defined as

$$\sigma_n(\mathbf{r}, \omega) = \frac{\sigma_{n,\mathrm{sample}}(\mathbf{r}, \omega)}{\sigma_{n,\mathrm{ref}}(\omega)} = s_n(\mathbf{r}, \omega) e^{i\phi_n(\mathbf{r},\omega)}, \qquad (2)$$

where $s_n$ and $\phi_n$ denote the referenced amplitude and phase, respectively.

In our study, we recorded hyperspectral datacubes on a polymer-blend sample consisting of polymethyl methacrylate (PMMA) domains embedded in a polystyrene (PS) matrix. The sample was prepared by drop casting a commercially available solution of PS and PMMA (Nanosurf) onto a silicon substrate, yielding a polymer film with an approximate thickness of 250 nm.

Figure 1(b) shows the amplitude and phase spectra acquired within a PMMA domain for the first three harmonic orders (*n*=1,2,3). The third-harmonic spectrum exhibits peaks at the characteristic PMMA vibrational resonance frequencies (indicated by vertical dashed lines) [39] and is nearly background-free, which explains why this harmonic is commonly used for nano-FTIR spectroscopy. Interestingly, the characteristic PMMA vibrational features are also discernible in the first and second harmonics, although with reduced contrast due to a pronounced wavy background. This background is characteristic of nano-FTIR spectroscopy and is also evident in the approach curves presented in the Supplementary Note 1. Thus, while the lower harmonics retain near-field spectroscopic information and offer a higher signal-to-noise ratio (SNR), this information is superimposed on a substantial background. The third harmonic, in contrast, provides much better background suppression but has a lower SNR. This illustrates the trade-off between background suppression and SNR across harmonic orders. Notice that higher noise can produce apparent but erroneous spectral features. For instance, in Fig. 1(d), the fluctuations in the third harmonic phase around $1050\ \mathrm{cm}^{-1}$ and $1790\ \mathrm{cm}^{-1}$ can be mistaken by peaks that do not correspond to any PMMA vibrational resonances.

By raster-scanning the sample over an $N_x \times N_y$ spatial grid, we acquired a hyperspectral datacube, whose amplitude and phase are shown in Figs. 1(c,d), respectively. The upper surfaces correspond to the spatial amplitude and phase maps at $1736\ \mathrm{cm}^{-1}$, close to the PMMA C=O stretching resonance. At the third harmonic, the PMMA domains and their nanoscale morphology are resolved with high spatial resolution, but the images exhibit pronounced noise due to the low SNR. In contrast, the first- and second-harmonic images are considerably less noisy and resolve the same nanoscale features despite the substantial background contributions. This suggests that the near-field signal in the lower harmonics has a substantially higher SNR than that in the third harmonic, even though it is superimposed on a significant background. Nevertheless, the lower signal harmonics are generally discarded in nano-FTIR spectroscopy because their near-field contributions cannot be readily separated from the substantial background.

## Cross-harmonic neural network method

IR hyperspectral cubes are high dimensionality data sets and thus amenable for denoising using machine learning approaches [50–52]. However, so far, DL methods are only beginning to be explored in s-SNOM [53–56], and while their potential for denoising is apparent [57], current DL protocols still operate on isolated single harmonic data, thus not making full use of the multiple signal harmonics of s-SNOM.

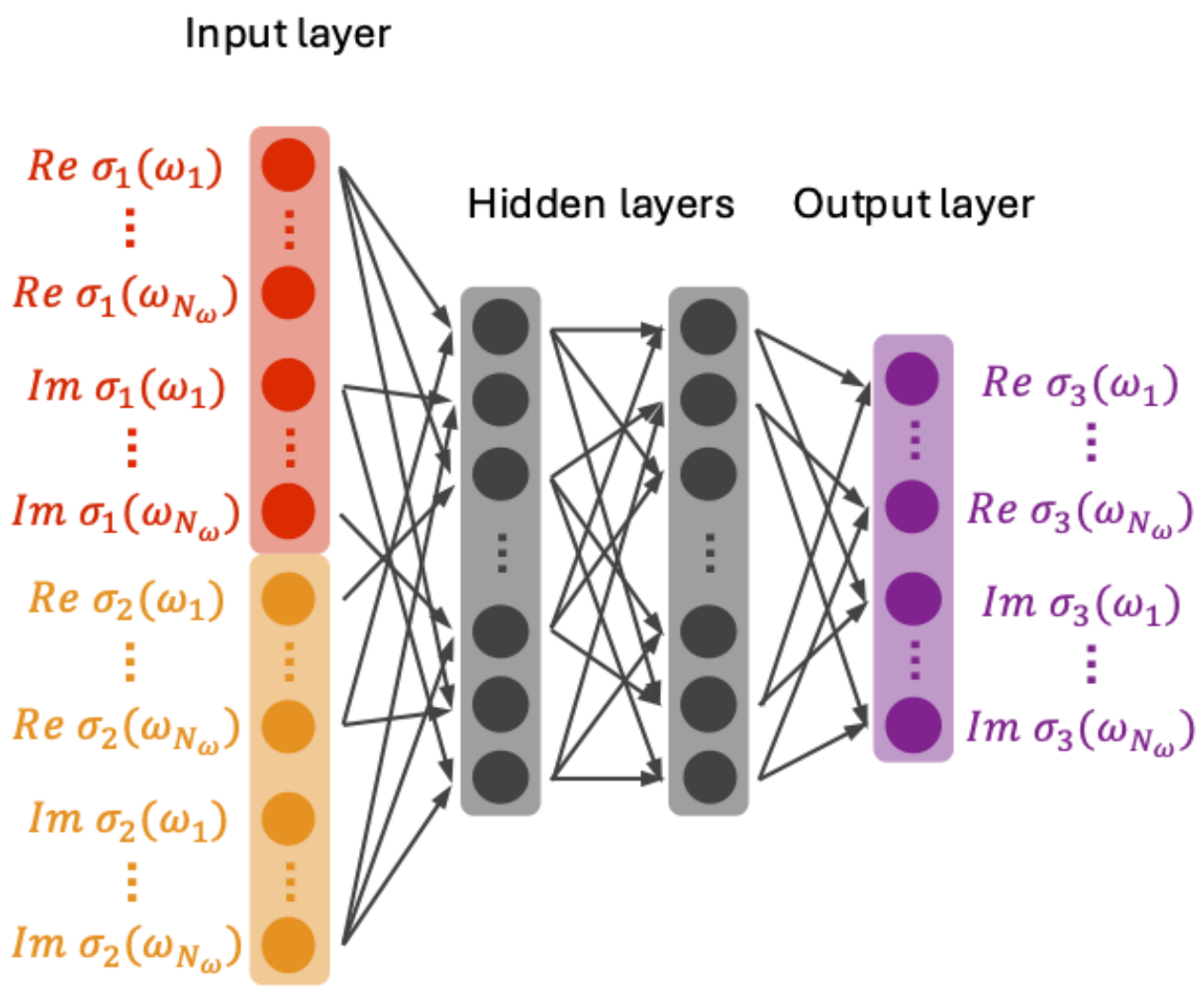

**Figure 2: Architecture of the CHNN.** For each spatial position $\mathbf{r}$, the neural network takes as input a one-dimensional vector formed by concatenating the real and imaginary spectra of the first and second harmonics, $\mathrm{Re}\sigma_1(\omega)$, $\mathrm{Im}\sigma_1(\omega)$, $\mathrm{Re}\sigma_2(\omega)$ and $\mathrm{Im}\sigma_2(\omega)$, each evaluated over all $N_\omega$ sampled wavenumbers. The neural network maps this input through a sequence of hidden layers, combining trainable linear transformations with nonlinear activation functions, onto the corresponding third-harmonic spectrum, represented by the concatenated real and imaginary components $\mathrm{Re}\sigma_3(\omega)$ and $\mathrm{Im}\sigma_3(\omega)$.

To exploit rather than discard near-field information contained in the background-contaminated lower signal harmonics, we introduce the cross-harmonic neural network (CHNN). As illustrated in Fig. 2, the CHNN is a multilayer perceptron that maps the complex-valued spectra $\sigma_n(\mathbf{r},\omega)$ of selected lower harmonics onto the corresponding spectrum of a selected higher harmonic at each spatial position $\mathbf{r}$. The network operates on complete spectra, simultaneously processing the values at all $N_\omega$ wavenumbers. Since the real-valued neural network used here cannot directly process complex-valued inputs, each complex-valued spectrum is represented by concatenating its real and imaginary parts into a single vector. In this work, we focus on mapping the first- and second-harmonic spectra to the third-harmonic spectrum. Other harmonic combinations were also investigated (see Supplementary Note 2). The third harmonic is chosen as the target because, as discussed above, it provides substantially better background suppression.

The resulting input is a one-dimensional vector of size $4N_\omega$, combining the real and imaginary parts of the first- and second-harmonic spectra. The output is a one-dimensional vector of size $2N_\omega$, containing the real and imaginary parts of the third-harmonic spectrum. The CHNN is trained using the measured hyperspectral data cubes shown in Fig. 1, with each spatial position $\mathbf{r}$ defining one training sample. Further details on hyperparameter selection, training, validation, and data treatment are provided in the Methods section.

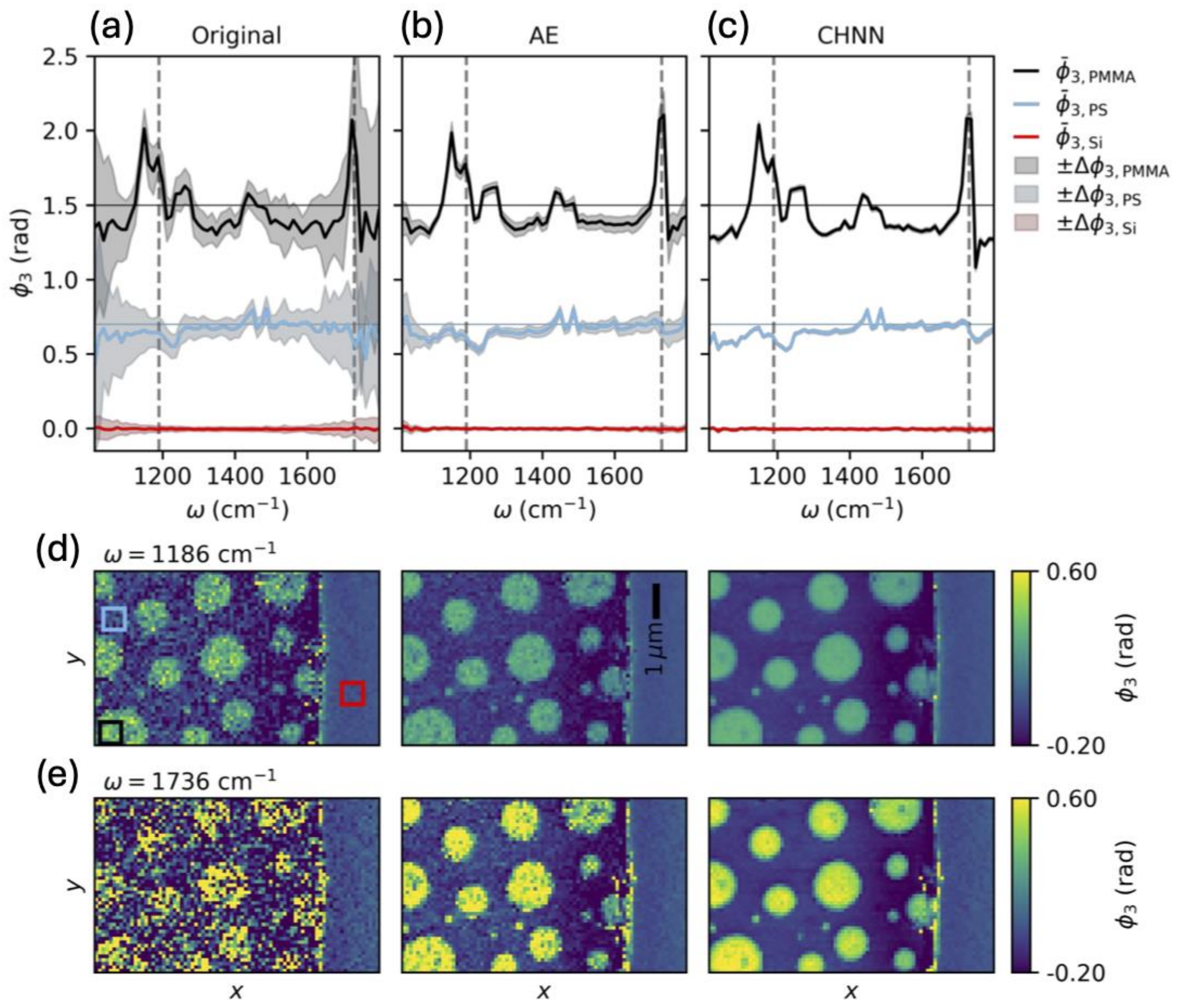

**Figure 3: Hyperspectral data cube reconstruction with the CHNN method. (a)** Phase spectra $\phi_3(\omega)$ obtained from ROIs containing pure PMMA, PS, and Si in the original hyperspectral data. PMMA, PS, and Si spectra are shown in black, blue, and red, respectively. The solid lines indicate the mean spectra, while the shaded areas indicate the corresponding mean $\pm$ one standard deviation across spatial positions within each ROI. **(b)** Phase spectra after denoising using the AE model. **(c)** Phase spectra obtained using the proposed CHNN method, which leverages the first two harmonics to reconstruct the third harmonic. **(d)** Left: experimental spatial map of the third-harmonic phase $\phi_3(\mathbf{r},\omega)$ at the representative wavenumber $1186\,\mathrm{cm}^{-1}$ indicated by a dashed vertical line in panel (a). Center: corresponding map obtained after denoising with the AE. Right: reconstruction of the phase map using a CHNN model, illustrating the improved spatial contrast and noise suppression achieved by the proposed cross-harmonic approach. **(e)** The same as in (d) but for wavenumber $1736\,\mathrm{cm}^{-1}$. The square patches in (d) indicate the ROIs used to compute the spectral statistics shown in panels (a)-(c).

## Enhanced hyperspectral nano-FTIR imaging

We evaluate the ability of the CHNN to improve the SNR of third-harmonic hyperspectral datacubes by denoising the third-harmonic spectra from the corresponding first- and second-harmonic spectra. Its performance is benchmarked against a conventional deep-learning (DL) denoising approach, the autoencoder (AE) [58,59]. An AE is an encoder-decoder neural network trained to reconstruct its input through a lower-dimensional latent representation. This bottleneck constrains the network to retain the dominant features of the input data while suppressing noise. Details of the AE architecture and implementation are provided in the Methods section.

In the following, we compare the phase spectra of the original, AE-denoised, and CHNN-reconstructed third-harmonic hyperspectral data cubes. The phase is obtained from the corresponding complex-valued data and is closely related to the local absorption of the sample, making it particularly suitable for chemical compositional mapping. Figs. 3(a–c) show the phase spectra from three homogeneous 7 × 7-pixel regions of interest (ROIs), each containing a single material: PMMA, PS, or Si. The ROI locations are indicated by squares in Fig. 3(d). Solid curves represent the mean spectrum within each ROI, while shaded regions indicate the standard deviation (Eqs. 3 and 4 in Methods). In addition, we analyze spatial maps extracted from the same phase cubes at 1186 and 1736 $cm^{-1}$, close to the PMMA C–O–C bending and C=O stretching resonances, respectively (Figs. 3(d,e)). These wavenumbers also represent different SNR conditions, with higher laser power and a stronger third-harmonic signal at 1186 $cm^{-1}$ than at 1736 $cm^{-1}$.

Both methods exhibit a clear improvement in SNR relative to the original third-harmonic data with the CHNN consistently outperforming the AE. This is evident in Figs. 3(a-c), where the standard deviation within each material ROI (PMMA, PS, and Si) decreases from the original data to the AE reconstruction and further to the CHNN reconstruction, indicating a substantial reduction in spectral variability and noise. The same trend is observed in the spatial maps of Figs. 3(d,e), where pixel-to-pixel fluctuations are reduced after denoising, with the CHNN showing the lowest residual noise. Importantly, the characteristic material contrast is preserved at both wavenumbers, with the PMMA domains (disks) exhibiting higher phase values than the surrounding PS matrix. This is consistent with the stronger absorption of PMMA at the selected wavenumbers.

As discussed above, the laser power varies across the measured spectral range, giving rise to different experimental SNR regimes: the central region, approximately $1100\text{-}1650\ \mathrm{cm}^{-1}$, is characterized by relatively high laser power, whereas the spectral edges, $1010\text{-}1100\ \mathrm{cm}^{-1}$ and $1650\text{-}1800\ \mathrm{cm}^{-1}$, correspond to lower-power regions and therefore higher noise levels. This difference is clearly reflected in Fig. 3(a), where the shaded regions are noticeably broader at the spectral edges. Remarkably, the CHNN reconstruction yields a similarly low residual noise level across both regimes. This behavior is illustrated by the two representative cases in Figs. 3(d,e): although the original third-harmonic image at $1736\ \mathrm{cm}^{-1}$, corresponding to the lower-power regime, is substantially noisier than that at $1186\ \mathrm{cm}^{-1}$, the CHNN reconstructions exhibit similarly low residual noise at both spectral positions. To quantify this behavior, we average the standard deviation within each ROI over the high-power central region and, separately, over the low-power spectral tails. For PMMA, for example, the averaged standard deviation of the original third-harmonic data increases markedly from $0.13\ \mathrm{rad}$ in the central region to $0.47\ \mathrm{rad}$ at the spectral edges. In contrast, the CHNN reconstruction yields an averaged standard deviation of approximately $0.02\ \mathrm{rad}$ in both regimes, despite the very different noise levels of the original data. The corresponding quantitative results for the remaining materials are provided in the Supplementary Note 2.

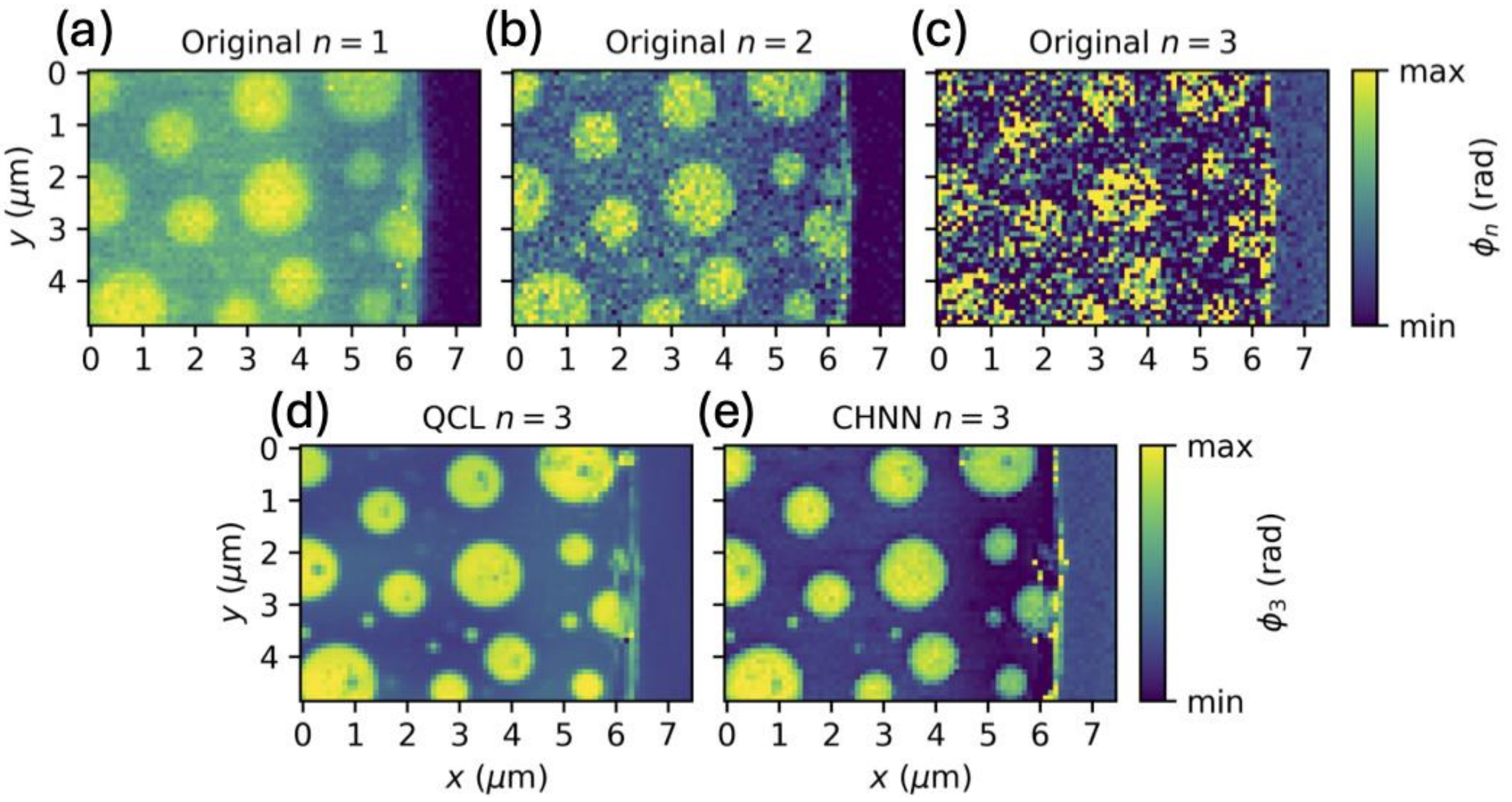

**Figure 4: Comparison between the original nano-FTIR phase images, an independent single-frequency QCL measurement, and the CHNN reconstruction close to the C=O stretching resonance of PMMA. (a-c)** Original nano-FTIR phase images acquired at $1736\ \mathrm{cm}^{-1}$ for the first, second, and third harmonic orders, respectively. **(d)** Independent QCL phase image acquired at $1740\ \mathrm{cm}^{-1}$. **(e)** CHNN reconstruction of the third-harmonic nano-FTIR phase image at $1736\ \mathrm{cm}^{-1}$. All panels show the same spatial region of the sample. The same colormap is used in all panels, while each image is displayed using its own $(\mathrm{min}, \mathrm{max})$ phase range: $(0, 0.5)$ rad for (a), $(0, 0.7)$ rad for (b), $(-0.2, 0.6)$ rad for (c), $(-0.2, 0.8)$ rad for (c) and $(-0.2, 0.6)$ rad for (e).

To evaluate the spatial fidelity of the CHNN reconstruction, we compare it with an independent single-frequency QCL image acquired at the C=O stretching resonance of PMMA (see Methods section for experimental details). It provides an independent reference for assessing whether the sample morphology and spatial features are reliably recovered by the reconstruction method. Additional comparisons with QCL images acquired at other wavenumbers associated with characteristic PMMA vibrational resonances are provided in the Supplementary Note 3.

Figs. 4(d) shows the QCL image acquired at 1740 $\mathrm{cm}^{-1}$ and the CHNN denoised reconstruction of the third-harmonic nano-FTIR image at 1736 $\mathrm{cm}^{-1}$ (Fig. 4(e)), respectively. The corresponding original nano-FTIR images for harmonics n=1,2,3 are shown in Fig. 4(a-c) for reference. The CHNN denoised reconstruction reproduces the morphology observed in the QCL image with high fidelity. Beyond the large PMMA disks, several smaller PMMA disks distributed horizontally within $x \in [0, 4]$ μm and $y \in [3, 4]$ μm can be clearly identified in both images, although they are difficult to distinguish in the original nano-FTIR maps shown in panels 4(a-c). The denoised reconstruction also resolves finer spatial features within the larger PMMA domains. In particular, small voids that are visible in the QCL image but not in any of the original nano-FTIR harmonic images appear in the CHNN denoised reconstruction. Examples are located around $(x, y) \simeq (0, 2.3)$ μm, (1, 4.5) μm, and (5.5, 0.5) μm. The agreement with the independent QCL measurement supports the spatial fidelity of the CHNN reconstruction and shows that it recovers features obscured by background contributions or noise in the original nano-FTIR data.

## Conclusions

We have introduced the cross-harmonic neural network, a DL-based signal enhancement strategy specifically designed for hyperspectral nano-FTIR data. The method exploits the intrinsic multi-harmonic structure of nano-FTIR measurements by learning a mapping from selected lower harmonics to a higher-harmonic response; here, we have focused on reconstructing the third harmonic from the first- and second-harmonic spectra. Rather than discarding the lower harmonics because of their stronger background contribution, the CHNN takes advantage of their higher overall SNR to recover near-field information that is already present in these signals but is masked by background and cannot be directly isolated. We demonstrated the approach on a PMMA/PS polymer blend, showing that the reconstructed third-harmonic hyperspectral data exhibit a substantial increase in SNR while preserving the characteristic spatial features and material-specific spectral response. Moreover, the CHNN consistently outperforms the single-harmonic AE benchmark, demonstrating the advantage of exploiting correlations between different harmonic orders rather than denoising the higher harmonic independently.

Independent single-frequency QCL measurements were used to evaluate the spatial fidelity of the CHNN reconstruction. The comparison shows that the CHNN reliably recovers the morphology observed in the QCL measurements, including the position and shape of the main PMMA domains as well as smaller spatial features within and around them that are difficult to distinguish in the original nano-FTIR images. This agreement provides independent evidence that the CHNN preserves the underlying spatial information while substantially increasing the SNR of the reconstructed third-harmonic response. These results demonstrate that near-field information that is difficult to access directly in the background-dominated lower nano-FTIR harmonics, and strongly affected by noise at higher harmonics, can be recovered with the aid of DL. More broadly, this work opens a route toward leveraging DL to improve the quality of hyperspectral near-field imaging.

# Methods

## Hyperspectral IR nanoimaging

Hyperspectral infrared nanoimaging was done on a commercial s-SNOM (neaSNOM, Attocube Systems GmbH) using the nano-FTIR module. The beam from a broadband mid-IR continuum laser based on the difference frequency generation (using the "C" Setting in the neaSNOM software to cover a spectral range between 1000 and 1800 $\mathrm{cm}^{-1}$) was focused on the metallic tip (PPP-NCSTAu, Nanoworld) of an atomic force microscope (AFM). The tip-scattered light was analyzed with an asymmetric Fourier transform interferometer and detected using a mercury-cadmium-telluride detector. To remove background scattering from the tip shaft and the sample, the AFM tip was vibrated vertically at a frequency of $\Omega = 139$ kHz and an amplitude of 81 nm. The detector signal, $I_{\mathrm{det}}$, was then demodulated simultaneously at the fundamental, second and third harmonics, $n\Omega$ with $n = 1,2,3$. Near-field interferograms were obtained by recording the demodulated detector signals, $I_n(d)$, as a function of the reference mirror position, $d$. For hyperspectral nanoimaging, near-field interferograms were recorded at each pixel $(x, y)$ of a 2D area of the sample surface, yielding 80 x 49 pixels and spanning an area of 8.0 μm × 4.9 μm (100 nm pixel size). To keep total imaging time manageable (4 hours), fast interferogram recording was employed at the following settings: 2 ms integration time per interferogram point at 2048 points per interferogram (4 sec). An interferogram length of 400 μm was chosen to resolve the relevant IR molecular vibration bands of the polymer blend with clarity (corresponding to 12.5 $\mathrm{cm}^{-1}$ spectral resolution). Balanced detection was employed to improve the SNR. [60] After recording, near-field interferograms were Fourier transformed to obtain the raw near-field spectra and subsequently referenced to the near-field spectra on the substrate material using standard protocols, as described in Eq. (2) of the main text. This procedure yielded three referenced near-field hyperspectral data cubes, $\sigma_n(\mathbf{r}, \omega)$, at signal harmonics, $n = 1,2,3$, from a single hyperspectral imaging scan.

For comparison, single-frequency "QCL" nanoimaging was performed on the same s-SNOM system using a wavelength-tunable QCL for tip and sample illumination and a pseudoheterodyne interferometer for detection of the tip-scattered light [48]. To this end, the QCL laser was tuned to one of the IR molecular vibration bands of PMMA and the same area as in the hyperspectral data cube was scanned at 50 nm pixel size. For a direct comparison with the AE and CHNN reconstructions (having a pixel size of 100 nm), the single-frequency IR images were spatially resampled in postprocess to a 100 nm pixel size by bilinear interpolation.

## Noise analysis

For each harmonic order $(n)$, an ROI mean phase spectrum is calculated as

$$\bar{\phi}_{n,M}(\omega) = \frac{1}{N_{\mathrm{ROI}}} \sum_{\mathbf{r} \in \mathrm{ROI}} \phi_n(\mathbf{r}, \omega), \tag{3}$$

where $M = \mathrm{PMMA, PS, Si}$ indexes the material and $N_{\mathrm{ROI}}$ is the number of pixels within the ROI. The corresponding spatial standard deviation is given by

$$\Delta\phi_{n,M}(\omega) = \sqrt{\frac{1}{N_{\mathrm{ROI}}} \sum_{\mathbf{r} \in \mathrm{ROI}} \left[\phi_n(\mathbf{r}, \omega) - \bar{\phi}_{n,M}(\omega)\right]^2}, \tag{4}$$

The same procedure is applied to the ROI amplitude spectra $s_n(\omega)$.

## Deep learning implementation

The training and validation datasets are constructed directly from the measured nano-FTIR hyperspectral data cubes. Each spatial position $\mathbf{r}$ provides one sample containing the complete spectrum at that position, yielding $N_{\mathrm{pixels}} = N_x N_y = 80 \times 49 = 3920$ samples. Before training, the samples were randomly shuffled and divided into 80% training and 20% validation subsets.

Since the magnitude of the scattering signal decreases with increasing harmonic order, the data for each harmonic were independently rescaled to the interval [0,1] using min-max normalization. For each harmonic $n$, the minimum and maximum values were calculated jointly over the real and imaginary components of the corresponding hyperspectral cube, and both components were rescaled using these same values.

For the CHNN, each sample consists of an input vector obtained by concatenating the real and imaginary parts of the first- and second-harmonic spectra over all $N_\omega = 64$ wavenumbers, giving $n_{\text{in}} = 256$, and a target vector containing the real and imaginary parts of the corresponding third-harmonic spectrum, with $n_{\text{out}} = 128$. The network is a fully connected multilayer perceptron composed of five hidden dense layers containing 200, 100, 100, 100, and 200 neurons, respectively, with sigmoid activation functions. The output layer is also dense and contains $n_{\text{out}} = 128$ neurons. The model was trained by minimizing the mean squared error (MSE) between the predicted and measured real and imaginary components of the third-harmonic spectrum. Training was performed for 1000 epochs using the Adam optimizer, with a learning rate of $10^{-3}$ and a mini-batch size of 64.

For the AE benchmark, each sample contains only the measured third-harmonic spectrum, represented by the concatenated real and imaginary components over the same $N_\omega = 64$ wavenumbers. The input and output therefore both have size $n_{\text{in}} = n_{\text{out}} = 128$. The AE is composed of dense layers arranged in an encoder-decoder architecture. The encoder contains layers of 200 and 100 neurons followed by a 10-dimensional latent bottleneck, while the decoder mirrors this structure with dense layers of 100 and 200 neurons before the 128-neuron output layer. Sigmoid activation functions are used throughout the hidden layers. The AE was trained using the same 80% - 20% data split, min-max rescaling procedure, MSE loss, number of epochs, Adam optimizer, learning rate, and mini-batch size as the CHNN.

## Acknowledgement

This work was financially supported by the Spanish "Ministerio de Ciencia, Innovación y Universidades", MICIU/AEI/ 10.13039/501100011033, through grants PID2023-148359NB-C21, PID2023-148359NB-C22 (also funded by ERDF/EU), PID2024-156602NB-I00 (also funded by ERDF/EU), RYC2023-044188-I (also funded by ESF+), CEX2024-001491-S (DIPC), CEX2023-001286-S (INMA) and CEX2020-001038-M (NANOGUNE). We also acknowledge the Aragón Regional Government through project QMAD (E09_23R). Funded by the IKUR Strategy on behalf of the Basque Government's Department of Science, Universities and Innovation.

## Competing interests

R.H. is a co-founder of Neaspec GmbH, which now is a part of Attocube Systems GmbH, a company producing s-SNOM systems, such as the one used in this study. The remaining authors declare no competing interests.

## Supporting Information Available

Supplementary Note 1: Approach curves.
Supplementary Note 2: First- to third-harmonic cross harmonic neural network
Supplementary Note 3: QCL comparison at additional wavenumbers

# Supplementary Note 1: Approach curves

We present the approach curves for the first three harmonic orders to assess their different background contributions in the amplitude signal. These approach curves have been taken on the silicon substrate of the sample considered in the main text. Figure S1 shows the approach curves for the near-field amplitude signal, $s_n(z)$, for the harmonics $n = 1,2,3$, normalized to the signal $s_n(z = 0)$. The $n = 3$ near-field amplitude signal rapidly decays towards zero with increasing tip–sample distance, indicating the absence of significant background, as expected. In contrast, the $n = 1$ and $n = 2$ near-field amplitude signals retain a significant non-zero signal even for large tap-sample distances, indicating the presence of background. To estimate the relative background contribution for each harmonic, the tails of the amplitude curves between $z = 100$ and $z = 180$ nm are fitted linearly and extrapolated to $z = 0$. The resulting intercept provides an estimate of the background fraction relative to the total amplitude signal: 93% for $n = 1$, 38% for $n = 2$, and 2% for $n = 3$, which are the expected values for these typical s-SNOM parameters used here (107 nm tapping amplitude, gold tip, taken on Si) and in the main text. Therefore, these approach curves confirm the strong presence of background in the amplitude signal for lower harmonic orders.

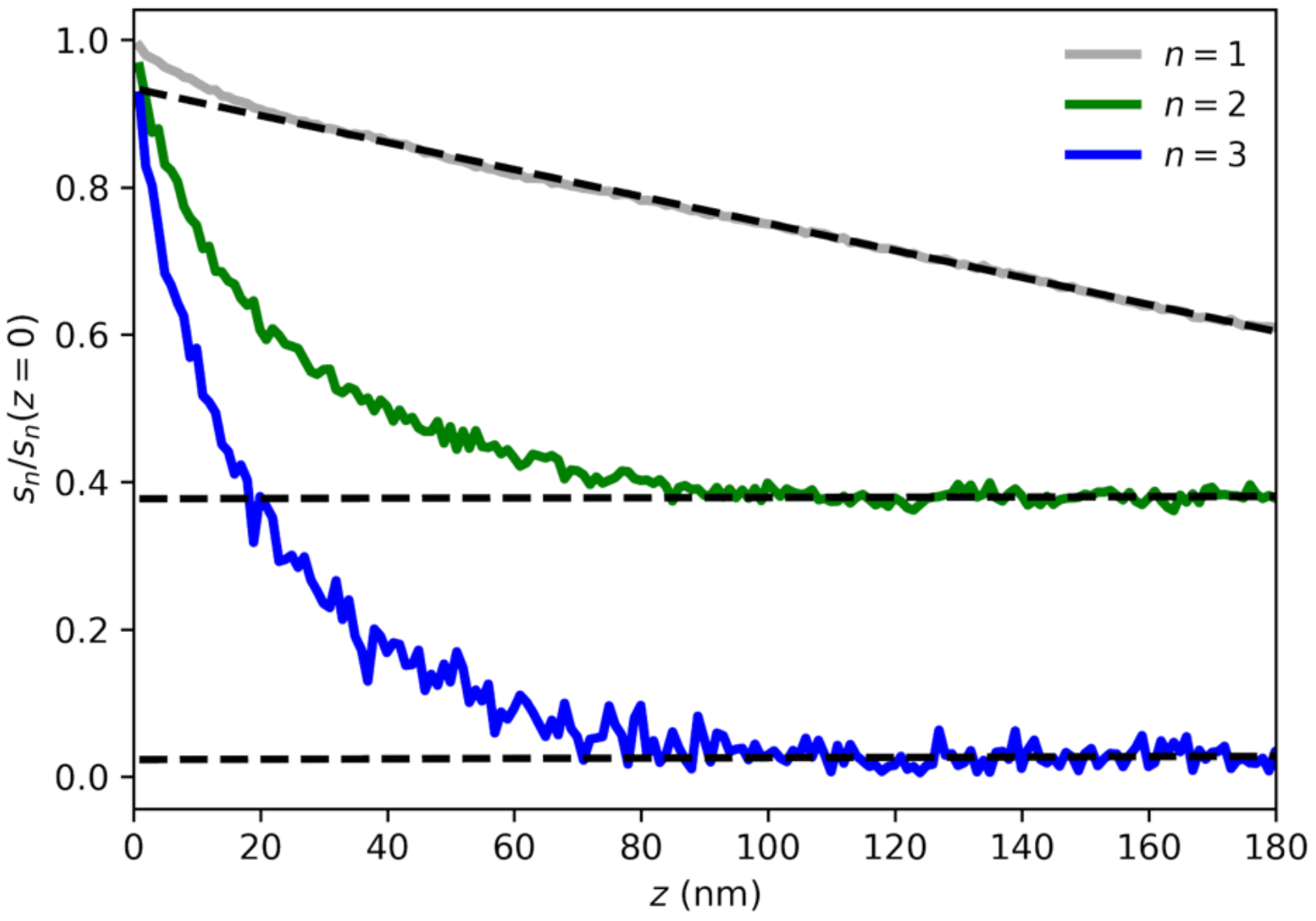

**Figure S1: Approach curves for the first three harmonic orders.** Near-field amplitude signal $s_n(z)$ ($n = 1,2,3$ in gray, green, and blue, respectively) measured on Si as a function of tip–sample distance $z$, with each harmonic signal normalized to its value at $z = 0$, under similar experimental conditions to those used for the nano-FTIR measurements in the main text. Dashed lines show linear fits to the long-distance tail (100 nm $< z <$ 180 nm). Their extrapolated intercepts at $z = 0$ provide an estimate of the background contribution relative to the total signal at contact.

## Supplementary Note 2: First- to third-harmonic cross-harmonic neural network (CHNN)

In this Supplementary Note, we present an alternative CHNN ($1 \rightarrow 3$) configuration that maps only the first-harmonic spectrum onto the third-harmonic spectrum, whereas in the main text the CHNN ($1,2 \rightarrow 3$) mapped both first and second spectra onto the third harmonic spectrum. We use the same hyperparameter configuration as that reported in the Methods section for the CHNN ($1,2 \rightarrow 3$) model. For direct comparison with the results presented in Fig. 3 of the main text, we repeat the same procedure to reconstruct the nano-FTIR hyperspectral data. Figure S2 shows, respectively, the original, AE-denoised, and CHNN ($1 \rightarrow 3$) reconstructed phase spectra obtained from three $7 \times 7$-pixel homogeneous ROIs, each containing a single material: PMMA, PS, or Si. On the one hand, the CHNN ($1 \rightarrow 3$) model substantially reduces noise in the same manner as in the CHNN ($1,2 \rightarrow 3$) model presented in the main text. Specifically, the averaged standard deviations for the CHNN ($1 \rightarrow 3$) model, together with those obtained for the CHNN ($1,2 \rightarrow 3$) and AE models, are reported in Table S1, for both central spectral region and edges.

On the other hand, this model does not recover the finer spatial features, such as the sub-300 nm PMMA particles in the PS matrix and the voids in the large PMMA domains. We attribute this to the larger probing volume of the first-harmonic signal, which contains less localized near-field information of the PMMA and more of surrounding PS in case of the small PMMA particles. In comparison, the CHNN ($1,2 \rightarrow 3$) model also uses the second harmonic as input, which has a smaller probing volume and probes the small spatial features with more sensitivity, therefore ensuring a correct reconstruction.

| | **Central region** (1100-1650 $cm^{-1}$) | | | | **Edges** (1010-1100 $cm^{-1}$ & 1650-1800 $cm^{-1}$) | | | |
|---|---|---|---|---|---|---|---|---|
| | **Original** | **AE** | **CHNN** $1,2 \rightarrow 3$ | **CHNN** $1 \rightarrow 3$ | **Original** | **AE** | **CHNN** $1,2 \rightarrow 3$ | **CHNN** $1 \rightarrow 3$ |
| **PMMA** | 0.1259 | 0.0432 | 0.0159 | 0.0142 | 0.4698 | 0.0786 | 0.0198 | 0.0188 |
| **PS** | 0.0942 | 0.0339 | 0.0135 | 0.0105 | 0.3383 | 0.0792 | 0.0150 | 0.0129 |
| **Si** | 0.0191 | 0.0098 | 0.0087 | 0.0089 | 0.0582 | 0.0191 | 0.0126 | 0.0146 |

**Table S1: Averaged standard deviation per material across the two spectral ranges defined in the main text.**

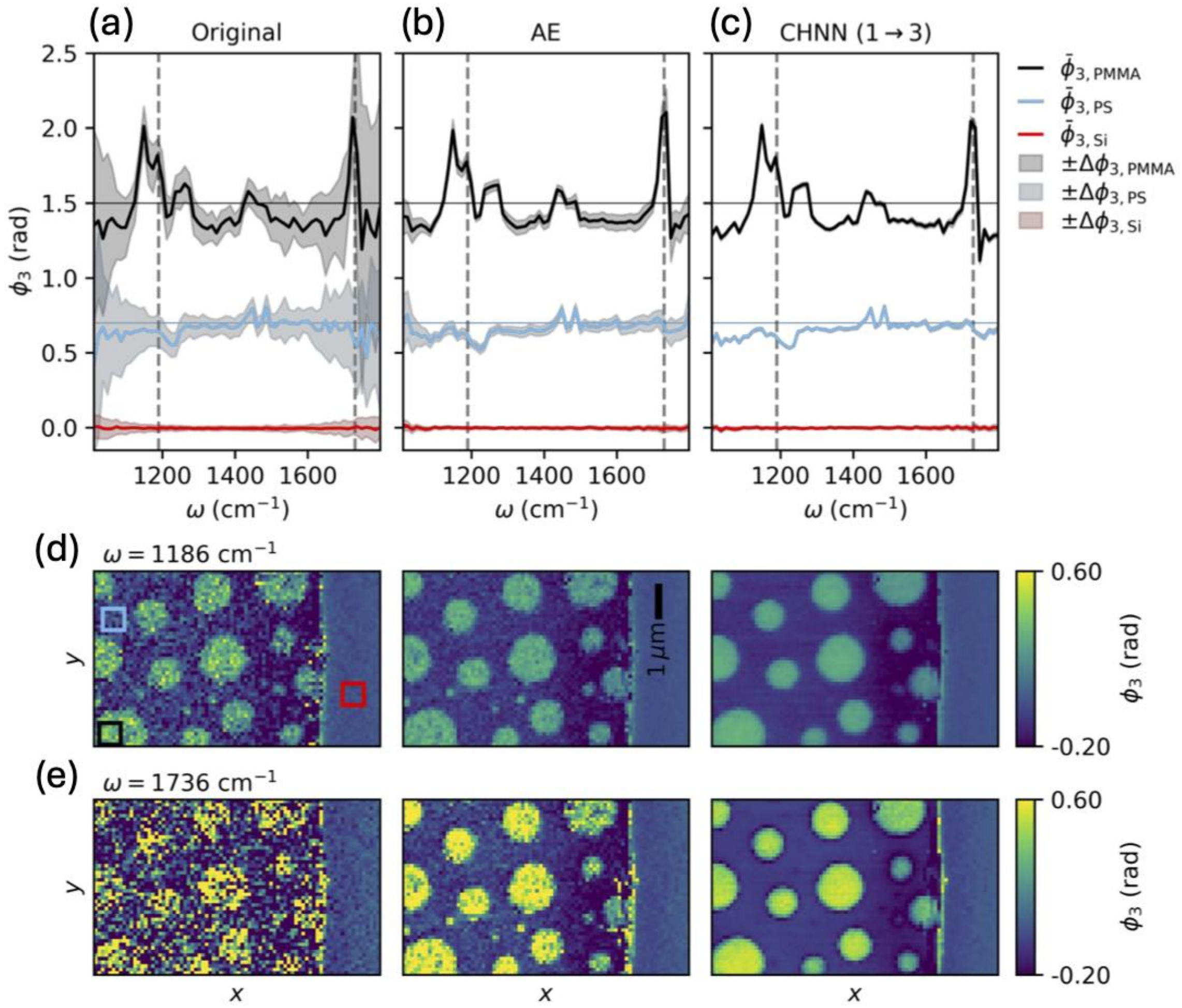

**Figure S2: Hyperspectral data cube reconstruction with the first- to third-harmonic CHNN. (a)** Phase spectra $\phi_3(\omega)$ obtained from ROIs containing pure PMMA, PS, and Si in the original hyperspectral data. PMMA, PS, and Si spectra are shown in black, blue, and red, respectively. The solid lines indicate the mean spectra, while the shaded areas indicate the corresponding mean $\pm$ one standard deviation across spatial positions within each ROI. **(b)** Phase spectra after denoising using the AE model. **(c)** Phase spectra obtained using the proposed first- to third-harmonic CHNN. **(d)** Left: experimental spatial map of the third-harmonic phase $\phi_3(\mathbf{r},\omega)$ at the representative wavenumber $1186\ \mathrm{cm}^{-1}$ indicated by a dashed vertical line in panel (a). Center: corresponding map obtained after denoising with the AE. Right: reconstruction of the phase map using a $1 \rightarrow 3$ CHNN model. **(e)** The same as in (d) but for wavenumber $1736\ \mathrm{cm}^{-1}$. The square patches in (d) indicate the ROIs used to compute the spectral statistics shown in panels (a)-(c).

## Supplementary Note 3: QCL comparison at additional wavenumbers

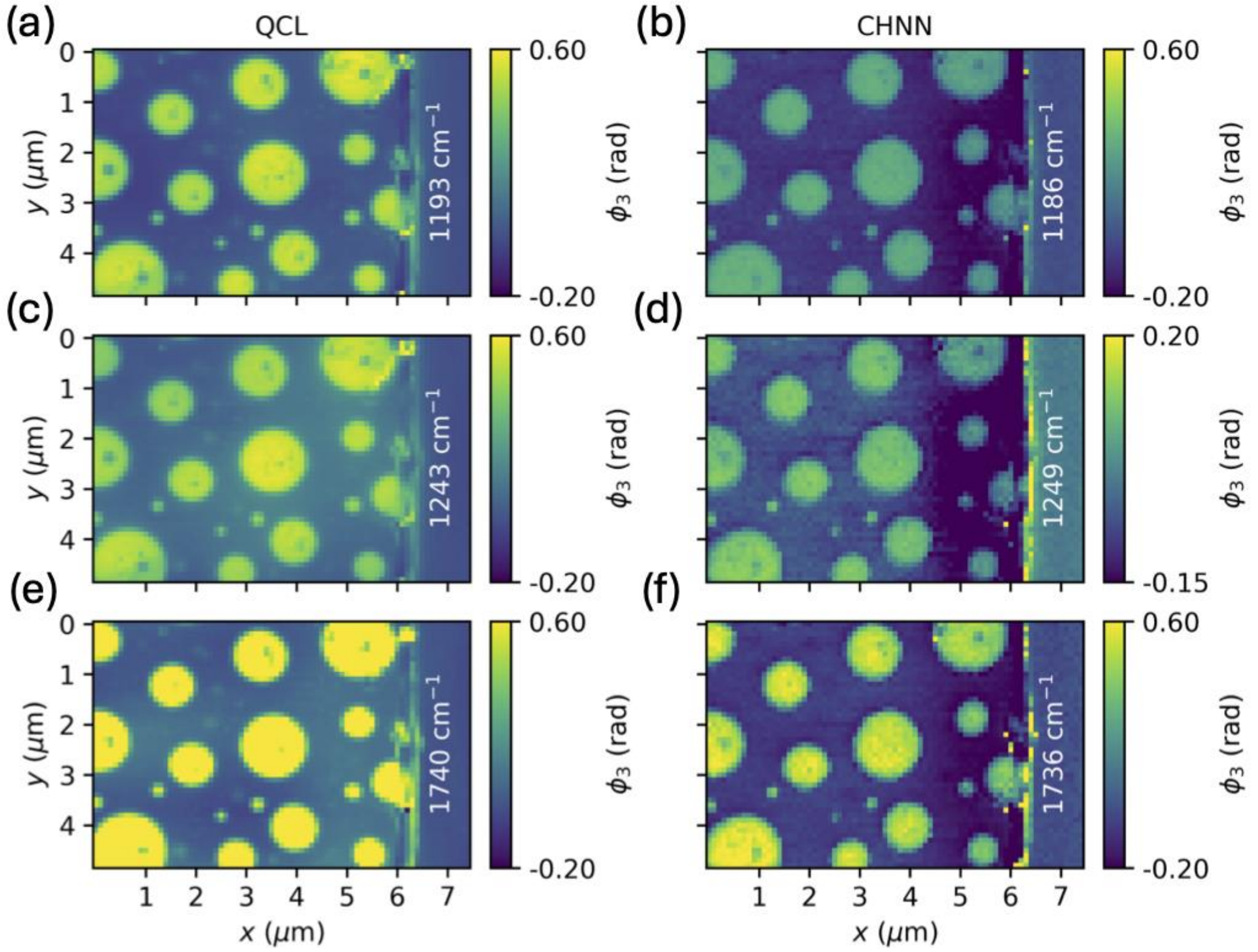

**Figure S3: Comparison between single-frequency QCL phase images and CHNN reconstructions at selected PMMA vibrational resonances. (a,b)** QCL phase image acquired at $1193\ \mathrm{cm}^{-1}$, close to the characteristic PMMA C–O–C bending resonance, and corresponding CHNN reconstruction at the closest sampled nano-FTIR wavenumber, $1186\ \mathrm{cm}^{-1}$. **(c,d)** QCL phase image acquired at $1243\ \mathrm{cm}^{-1}$, (close to the PMMA C–C–O stretching resonance, and corresponding CHNN reconstruction at the closest sampled nano-FTIR wavenumber, $1236\ \mathrm{cm}^{-1}$. **(e,f)** QCL phase image acquired at $1740\ \mathrm{cm}^{-1}$, close to the PMMA C=O stretching resonance, and corresponding CHNN reconstruction at the closest sampled nano-FTIR wavenumber, $1736\ \mathrm{cm}^{-1}$. In all panels, the same spatial region of the sample is shown.